\documentclass{article}
\usepackage{spconf,amsmath,graphicx,hyperref}

\title{PROFICIENCY-AWARE ADAPTATION AND DATA AUGMENTATION FOR ROBUST L2 ASR}
\name{Ling Sun$^{*}$\thanks{These authors contributed equally to this work.} \qquad
      Charlotte Zhu$^{*}$\footnotemark[1] \qquad
      Shuju Shi\thanks{Corresponding author: shi16@iu.edu}}
\address{Department of Linguistics, Indiana University, Bloomington, USA}
\begin{document}

\maketitle

\begin{abstract}
General-purpose ASR underperforms for atypical speakers, such as L2 learners, reinforcing bias and limiting use in education and accessibility. Using the CEFR-graded Speak \& Improve corpus, we show that naïve fine-tuning of Whisper reduces average WER but simultaneously widens disparities and disproportionately harms lower-level learners. To address this, we propose two strategies: (i) proficiency-aware multitask learning, jointly optimizing ASR with proficiency classification, and (ii) targeted augmentation, applying spectrogram masking to low-proficiency speech to counter imbalance. These approaches reduce WER by up to 29.4\% (relative) and insertion/deletion errors by as much as 58.6\% (relative). Crucially, despite the dataset’s severe imbalance reflecting real-world distributions, both strategies consistently narrow proficiency gaps, advancing equitable ASR for L2 learners.
\end{abstract}


\section{Introduction}
Automatic speech recognition (ASR) has been widely applied in voice assistants, accessibility technologies, and language learning platforms. However, general purpose ASR systems perform worse with atypical speakers, such as individuals with speech impairments and non-native (L2) speakers. This performance gap not only leads to systematic biases in real-world deployments \cite{chan2022training, mujtaba2024lost}, but also limits the applicability of recent advances in ASR to critical domains, including assistive healthcare and, more directly, language education. For language learners in particular, reliable ASR is crucial: prior studies show that ASR-guided feedback can accelerate L2 oral proficiency development \cite{cucchiarini2009oral}. Closing the gap for L2 speakers is therefore not only a technical challenge, but also an educational necessity.

A large body of L2 speech research documents structural characteristics that distinguish L2 from L1 speech, including segmental deviations (accents) \cite{tu2018investigating} and temporal disfluency (pauses, hesitations, repair) \cite{yan2025disfluency}. These features pose challenges for general-purpose ASR models, which are largely trained on L1-dominant corpora \cite{chan2022training}.

Among these challenges, accent variation has received the most attention. Early work improved model robustness with lexicon/pronunciation variance promotion \cite{goronzy2004generating}, acoustic-model adaptation \cite{bouselmi2007combined}, joint acoustic and accent  modeling/embedding \cite{yang2018joint, jain2018improved}, and domain adversarial training \cite{sun2018domain, na2021accented}. More recently, Mu et al. proposed a generative error-correction framework \cite{mu2025mixture}, where a fine-tuned LLM integrates speech embedding with word- and phoneme-level N-best outputs. Their model achieved substantial word error rate (WER) reductions on multi-accent English, outperforming strong baselines such as Whisper-large-v3.

Despite progress in accent-robust ASR, proficiency robustness remains a critical challenge for L2 speech. Importantly, proficiency variations are characterized by both segmental (e.g., mispronunciations) and temporal deviations (e.g., stress timing, pause durations). These patterns scale with learner proficiency. For instance, lower-proficiency learners show more frequent and longer pauses \cite{ williams2019pause}, which in turn influence listeners’ fluency judgments \cite{bosker2013makes}. However, widely used non-native English corpora such as L2-ARCTIC and AESRC-2020 do not include proficiency annotations \cite{shi2021accented,zhao2018l2}; in AESRC \cite{zhao2018l2}, results are reported by accent categories, not proficiency levels. This design limits our ability to analyze subgroup disparities and hinders adaptation to learners at different developmental stages. Moreover, advances in modeling disfluent L2 speech can directly benefit ASR on other disfluent speech types, such as stuttering \cite{mujtaba2024lost}.

This work addresses this gap by adapting speech foundation models with the Speak \& Improve Corpus, a large CEFR-graded L2 English speech dataset \cite{knill2024speak}. To our knowledge, this is the first work to systematically investigate proficiency-aware adaptation of foundation ASR models. Unlike accent-aware methods that focus on segmental deviations, our approach targets proficiency deviations that include both temporal and segmental disparities across learner groups.

Our contributions are threefold: (1) We provide the first systematic analysis of how ASR errors scale with CEFR proficiency, showing that performance disparities cannot be explained by data imbalance alone. (2) We demonstrate that naïve fine-tuning reduces average WER but worsens disparities, revealing the risks of proficiency-agnostic adaptation. (3) We propose two proficiency-aware strategies that reduce WER and narrow the gaps across groups: multitask learning with proficiency supervision, and targeted augmentation of low-proficiency data using spectrogram masking \cite{Park_2019} to mitigate imbalance.

\section{Methodology}
\subsection{Dataset}\label{dataset}
\begin{figure}[t]
\includegraphics[width=\columnwidth]{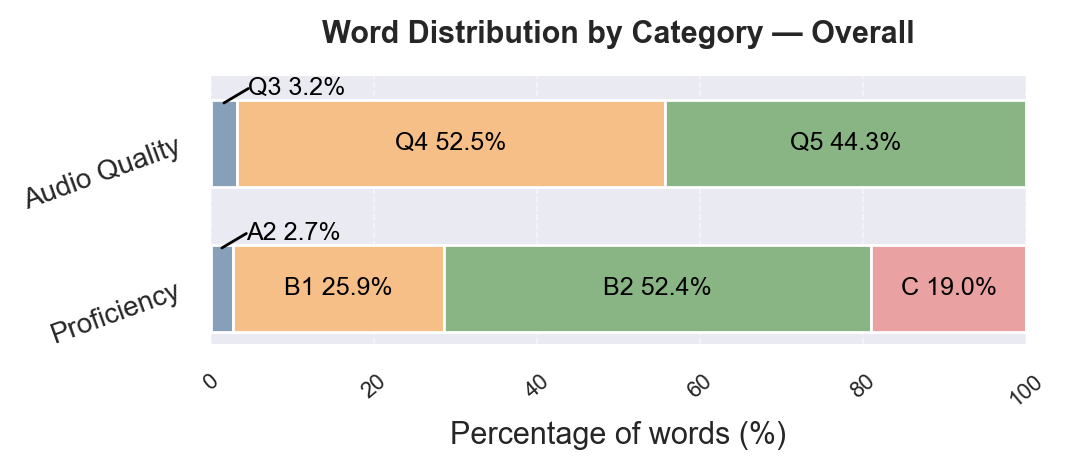}
  \caption{Word distribution in the S\&I corpus by label category.
    A2–C1 indicate CEFR proficiency levels (A2 = low, C1 = high).
    Q3–Q5 denote audio quality (Q3 = low, Q5 = high).}
  \label{fig:corpuspercentage}
\end{figure}

The dataset used in this study is the Speak \& Improve (S\&I) corpus, developed by the University of Cambridge in collaboration with Cambridge University Press \& Assessment (CUP\&A) and English Language iTutoring Ltd (ELiT) \cite{knill2024speak}. It comprises around 315 hours of L2 English learner speech, including 73.8 hours of manually transcribed data with error and disfluency annotations, partitioned into 28.2 hours for training, 22.9 hours for development, and 22.7 hours for evaluation. All recordings are graded according to the CEFR proficiency scale (A2–C1). In addition, the corpus provides metadata on audio quality ratings and task types.

The S\&I corpus shows clear imbalance across both audio quality ratings and proficiency levels, as in Figure~\ref{fig:corpuspercentage}, which closely mirrors real-world distributions. Among transcribed recordings, most are of higher audio quality (Q4–Q5), with as few as 3.2\% of words of low-quality (Q3). Proficiency distribution is likewise skewed, with B2 learners making up the largest share, while A2 and C1 are underrepresented. Notably, A2 data account for only 2.7\% of the words. Such imbalance presents challenges for training, as models risk overfitting to majority conditions and underperforming on lower-proficiency or lower-quality subsets.

\subsection{Model finetuning}
\begin{figure}[t]  \includegraphics[width=.9\columnwidth]{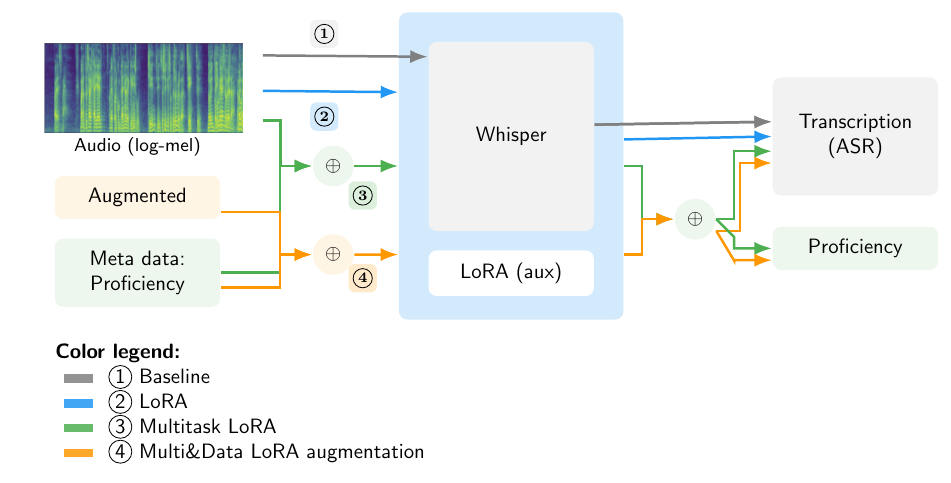}
  \caption{The pipelines used in this study, including the Whisper baseline and three mitigation strategies.}
  \label{fig:methodSummary}
\end{figure}

We adopt the Whisper-small model\footnote{\url{https://huggingface.co/docs/transformers/en/model_doc/whisper}} as our baseline for ASR performance on the S\&I corpus. We then design and evaluate three mitigation strategies (Fig~\ref{fig:methodSummary}): (i) \emph{LoRA adaptation}, to test whether parameter-efficient exposure to L2 speech alone suffices; (ii) \emph{multitask ASR with auxiliary  proficiency classification}, to explicitly model heterogeneous speech properties across proficiency levels; and (iii) \emph{targeted augmentation}, to counter class imbalance in the current corpus. 

\subsubsection{Naïve LoRA Adaptation}
We apply Low Rank Adaptation (LoRA) \cite{hu2021loralowrankadaptationlarge} to fine tune Whisper on the S\&I corpus. Adapters are inserted in selected projection matrices within attention and feed forward blocks, and only these small modules are updated. This uses the full L2 training set without correcting class imbalance or proficiency heterogeneity, giving a controlled baseline to test whether simple exposure improves recognition.

LoRA enables economic training and limits drift from the pretrained model, so gains can be attributed mainly to exposure. Motivated by evidence that listeners adapt to accented speech after limited exposure \cite{clarke2004rapid}, we expect this light adaptation to help Whisper accommodate L2 speech. We evaluate the naïve LoRA fintuned model on aggregate WER and on CEFR stratified metrics to quantify any remaining disparities.

\subsubsection{Multitask ASR \& Proficiency Classification}

Prior work shows that adding an auxiliary accent task can improve robustness by conditioning acoustic representations on accent variation \cite{yang2018joint, jain2018improved}. We adapt this idea to proficiency: starting from the LoRA setup, We extend Whisper by attaching a lightweight MLP classifier to the encoder, mean-pooling the final encoder states to predict CEFR proficiency alongside transcription.

Training is multitask: the ASR loss $\mathcal{L}_{\text{ASR}}$ is the standard Whisper sequence objective, and the proficiency classifier loss $\mathcal{L}_{\text{CLS}}$ is cross-entropy. 
The objective is a weighted sum \eqref{loss}, where both losses update the encoder and LoRA adapters, while the decoder is optimized only with $\mathcal{L}_{\text{ASR}}$.

\begin{equation}
\mathcal{L} \;=\; \lambda_{1}\,\mathcal{L}_{\text{ASR}} \;+\; \lambda_{2}\,\mathcal{L}_{\text{CLS}},
\quad \lambda_{1}=0.9,\;\lambda_{2}=0.1.\label{loss}
\end{equation}

\subsubsection{Data Augmentation}
The scarcity of A2 speech in the current dataset calls for data augmentation. However, our augmentation design is constrained by the need to preserve the validity of the proficiency label. Many common signal transforms such as speed perturbation can alter (supra)segmental realization, which are cues for proficiency; applying them risks corrupting the CEFR proficiency label that is pivotal for our task. We therefore adopt spectrogram augmentation (SpecAug) \cite{Park_2019}, which masks short contiguous spans in time and frequency on log mel features while leaving global proficiency intact. Augmentation is applied to A2 speech only to both counter class imbalance and adds local variability without changing the underlying proficiency label.

To isolate the contribution of augmentation, we first evaluate a pure augmentation setting that adds SpecAug to the LoRA baseline while keeping the optimizer, sampling, and loss weights fixed. We then combine the same SpecAug schedule with the multitask proficiency objective to test complementarity. At inference time augmentation is disabled. In ablations we report aggregate WER and CEFR stratified metrics.


\section{Results \& Discussion}

Table~\ref{tab:final_eval_oldloss} presents WER results across proficiency groups on the evaluation set. Relative to the baseline (WER of 10.2\%), LoRA reduces errors to 9.2\%, and the multi-task setup yields further improvement (8.1\%). Data augmentation alone achieves 7.4\%, while combining multi-task learning with augmentation provides the best performance at 7.2\%. Overall, all fine-tuned models substantially outperform the baseline, with Multi+Data achieving the best results, corresponding to a 29.4\% relative reduction in WER. All improvements over the baseline are statistically significant ($p_\text{boot}<$0.05).\footnote{We performed paired bootstrap over utterances ($B=10^{4}$). Let \textit{BL} and \textit{PR} denote the baseline and proposed systems. Each resample draws $N$ utterances with replacement, preserving BL/PR pairing. $\Delta\mathrm{WER}=\frac{1}{N}\sum_{u}\big[\mathrm{WER}_{\mathrm{PR}}(u)-\mathrm{WER}_{\mathrm{BL}}(u)\big]$. The 95\% percentile CI and a two-sided bootstrap $p$-value $2\min\{\Pr(\Delta\le0),\Pr(\Delta\ge0)\}$ are reported.}

\begin{table}[t]
\centering
\caption{WER over the evaluation dataset for all models, with the best results highlighted in bold.} 
\resizebox{\linewidth}{!}{
\begin{tabular}{ccccc}
\hline
Baseline & LoRA & Multi & DataAug & Multi+Data  \\
\hline
10.2  & 9.2 & 8.1 & 7.4 & \textbf{7.2}\\
\hline
\end{tabular}
}
\label{tab:final_eval_oldloss}
\end{table}

\begin{figure*}[t]
  \centering
\includegraphics[width=\textwidth]{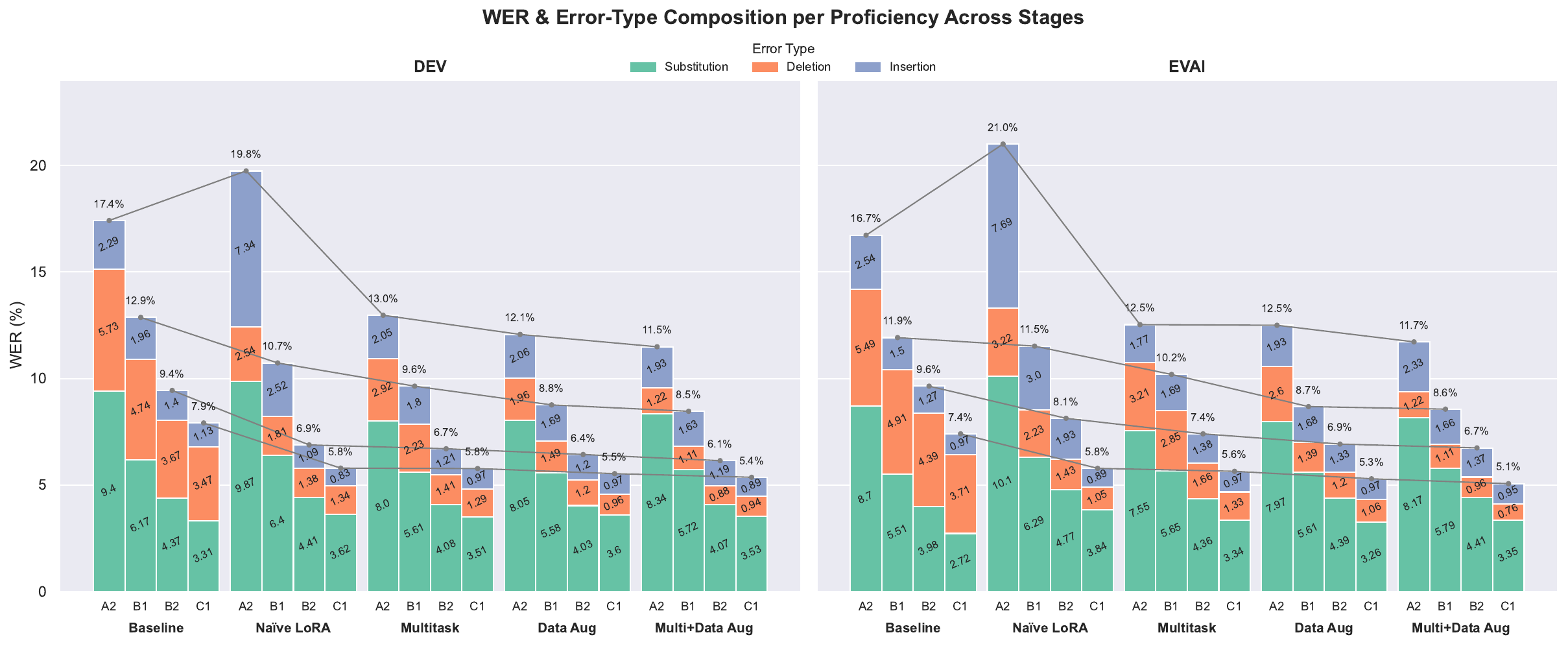}
  \caption{WER breakdown by error type over the development and evaluation dataset across five systems: baseline Whisper-small, LoRA fine-tuning, LoRA with auxiliary proficiency head, LoRA with data augmentation, and LoRA with both multitask learning and augmentation.}
  \label{fig:resultsSummary}
\end{figure*}

\subsection{ASR performance scales with proficiency}\label{error} 

Fig~\ref{fig:resultsSummary} reports WER broken down by error type across CEFR proficiency groups in both development and evaluation set. In the evaluation set, across all five systems, a clear trend emerges: ASR performance scales with proficiency level. The lowest-proficiency speakers (A2) consistently yield the highest WER, whereas performance improves steadily from B1 to C1. For instance, the baseline Whisper-small renders nearly 17\% WER on A2 speech but under 8\% on C1.

A natural concern is that the observed trend in the fine-tuned models may simply reflect data imbalance, as A2 constitutes only a small portion of the dataset (Fig~\ref{fig:corpuspercentage}). However, C1 speech is also relatively scarce (16.5\%) yet yields the best recognition performance across models. This asymmetry indicates that the observed gradient is not a mere artifact of data availability, but rather reflects systematic differences in speech characteristics across proficiency levels. Recognition improves as proficiency increases, as higher-proficiency speech more closely resembles the dominant L1-like training data. In other words, proficiency is a key latent variable in L2 ASR, calling for proficiency-aware training and evaluation.

Error composition further supports this point. For the baseline Whisper-small, overall substitutions, which frequently arises from accent-induced deviations, constitute less than half of total errors (46\%) , with insertions and deletions making up the rest (Fig~\ref{fig:resultsSummary}). This indicates that temporal fluency, such as hesitations, pauses, and filler words characteristic of lower-proficiency speech, plays an equally important role in recognition errors. Thus, L2 speech recognition requires attention to both segmental and time-sensitive variations, both of which reflected by proficiency labels.

\subsection{Naïve LoRA fine-tuning reinforces disparity} 

The proficiency-dependent performance patterns observed above suggest that adaptation strategies must account for learner variation across different proficiency groups. Indeed, when naïve LoRA fine-tuning is applied without proficiency awareness, the gains are unevenly distributed across groups.
As shown in Fig.~\ref{fig:resultsSummary}, for the naïve LoRA model,  higher-proficiency speakers (B2–C1) see consistent gains ($\Delta>$0, paired sign test $p<$0.01), but A2 performance significantly worsens ($\Delta<$0, paired sign test $p<$0.01), with WER increasing by a relative 20\% on the development set and 21\% on the evaluation set. This exposes a key risk: proficiency-agnostic adaptation can reduce general error rates while exacerbating disparities for the very learners most in need of ASR support.

Error analysis shows that the A2 degradation in the development set is driven primarily by insertion errors. Among the 295 insertions on A2 speech, the most frequent are short function words such as ``and'' (224), ``a'' (8), and ``the'' (6). This pattern suggests that the fine-tuned model overfits to filler-like usage in disfluent speech, amplifying errors for lower-proficiency speakers. These findings underscore that speech differs systematically across proficiency groups and that proficiency-agnostic adaptation conflates these distributions, leading to suboptimal and inequitable predictions.

\subsection{Proficiency-aware multitasking and data augmentation reduce disparities in time-sensitive errors}

In contrast, both of our proficiency-aware adaptation strategies yield consistent improvements across all proficiency groups, providing significant gains for lower proficiency speech (A2, B1). On the evaluation set, all adaptation strategies significantly reduced WER relative to the Naïve LoRA baseline ($p_\text{boot}<$0.05). Multitask learning achieved significant reduction at A2 ($\Delta$=0.0556, $p_\text{boot}<$0.01) and B1 ($\Delta$=0.0119, $p_\text{boot}$=0.05). Data augmentation yielded larger overall improvements, lowering sum/ave WER to 7.4\%, with significant gains across A2, B1, and B2 ($p_\text{boot}<$0.05). The combined model with multitask learning and data augmentation further narrows the residual proficiency gap (A2: $\Delta$=0.0611, $p_\text{boot}<$0.001; B1: $\Delta$=0.0239, $p_\text{boot}<$0.001), delivering the most equitable outcome and the most competitive sum/ave WER of 7.2\%.

Moreover, multitask and data augmentation more effectively mitigates the temporal disfluency errors, i.e., insertions and deletions. The mitigation strategies suppress the spiking insertion error rate most detrimental to A2 speech in the naïve LoRA model (7.34\% for development and 7.69\% for evaluation). They reduce insertion error to $\sim$2\%, and deletions fall from 5.49\% in the baseline to a similar floor. This is expected: CEFR proficiency labels are accent-agnostic, so proficiency-aware models and low-proficiency-focused augmentation are naturally better positioned to address proficiency-related variation, which primarily manifests more as temporal disfluencies (pauses, hesitations, and filler overuse) than segmental accent differences \cite{bosker2013makes,derwing1997accent}.

Taken together, these results demonstrate that naïve adaptation to L2 speech can worsen disparities, but proficiency-aware multitask learning and augmentation effectively suppress time-sensitive error modes and restore balance across proficiency groups. This highlights the importance of proficiency awareness for both accuracy and fairness in L2 ASR.

\section{Conclusion}
Our experiments reveal three key findings. First, proficiency is a critical latent variable in L2 ASR, and ignoring it mischaracterizes system performance. Second, naïve fine-tuning exposes the risk of proficiency-agnostic adaptation: it reduces average WER but exacerbates inequality. Third, proficiency-aware multitask learning and targeted augmentation directly address the spiking insertions that disproportionately harm low-proficiency speakers, yielding both lower WER and more equitable outcomes across proficiency groups. These results highlight proficiency as a crucial dimension for fair and effective L2 ASR.

Future work will focus on improving the proficiency classifier, whose F1 score remains limited by class imbalance, through more robust balancing strategies. In addition, since substitutions remain a dominant segment-sensitive error, we plan to explore accent-robust modeling approaches to further enhance performance.


\bibliographystyle{IEEEbib}

\end{document}